\documentstyle[prl,twocolumn,aps,floats,epsfig]{revtex}

\begin{document}

\draft
\twocolumn[\hsize\textwidth\columnwidth\hsize\csname
@twocolumnfalse\endcsname

\title{Seven energy/temperature scales in hole-doped cuprates}
\author{A Mourachkine}
\address{Free University of Brussels, CP-232, 
Blvd du Triomphe, B-1050 Brussels, Belgium} 

\date{Received}
\maketitle

\begin{abstract}
{\bf Abstract} The purpose of this paper is to discuss a phase diagram of 
hole-doped cuprates. 
Hole-doped cuprates have a very rich phase diagram as a function of doping. 
This is due to a charge inhomogeneity in CuO$_2$ planes: the charge 
distribution in CuO$_2$ planes is inhomogeneous both on a nanoscale and 
macroscale. Depending on the doping level, the CuO$_2$ planes comprise 
at least four different types of clusters having a size of a few nanometers. 
Each type of clusters has its own features. As a consequence, the common 
phase diagram consists of, at least, seven different energy/temperature 
scales. As an example, we consider a phase diagram of 
Bi$_{2}$Sr$_{2}$CaCu$_{2}$O$_{8+x}$. 
\end{abstract}

]

\section{Introduction} 

The phenomenon of high-$T_c$ superconductivity was discovered in cuprates 
[1]. The understanding of a phase diagram of cuprates is important since it 
is directly related to the understanding of the physics of high-$T_c$ 
superconductors. 

In the literature, one can find a few different phase diagrams for 
superconducting cuprates. As an example, figure 1 shows four different 
phase diagrams of hole-doped cuprates, taken from the literature. All 
four phase diagrams show the commensurate antiferromagnetic phase at 
low doping level $p$ and the superconducting phase with a transition 
temperature $T_c$. These two phases are long-ranged. The commensurate 
antiferromagnetic phase appears at a N\'eel temperature $T_N$. If in 
La$_{2-x}$Sr$_x$CuO$_4$ (LSCO), these two phases are separated by some 
distance, in YBa$_2$Cu$_3$O$_{6+x}$ (YBCO) the distance between them is 
almost absent. In spite of this and some other small differences, there is 
good agreement on the existence of these two phases in the phase diagram 
of superconducting cuprates. In figure 1, one can see that the disagreement 
among all these phase diagrams is mainly in the dispositions of the 
temperature scales situated above the superconducting phase. In other words, 
there is no consensus on the physics of normal state of cuprates.

Soon after the discovery of high-$T_c$ superconductivity it became clear that 
the understanding of normal-state properties of cuprates is crucial for the 
understanding of the mechanism of high-$T_{c}$ superconductivity. Indeed, 
there is an interesting contrast between the development of the physics of 
cuprates and that of the physics of conventional superconductors. Just 
before the creation of the BCS theory, the normal-state properties of 
conventional metals were very well understood; however superconductivity 
was not. The situation with the cuprates was just the opposite: at the time 
high-$T_{c}$ superconductivity was discovered, there already existed a good 
understanding of the phenomenon of superconductivity, but the normal-state 
properties of the cuprates were practically unknown. What is interesting is 
that 16 years after the discovery [1] their
normal-state properties are still  ``unknown.'' The problem is not that there
is no explanation of these unusual  properties, the problem is that there are
too many of them. 

There is good agreement on the existence of the so-called pseudogap which
occur at $T^{\ast} > T_c$. There is {\em even} consensus on the doping
dependence  of pseudogap in hole-doped cuprates: the magnitude of
pseudogap decreases as the doping level increases. 
\begin{figure}[h]
\leftskip-10pt
\epsfxsize=0.55\columnwidth
\centerline{\epsffile{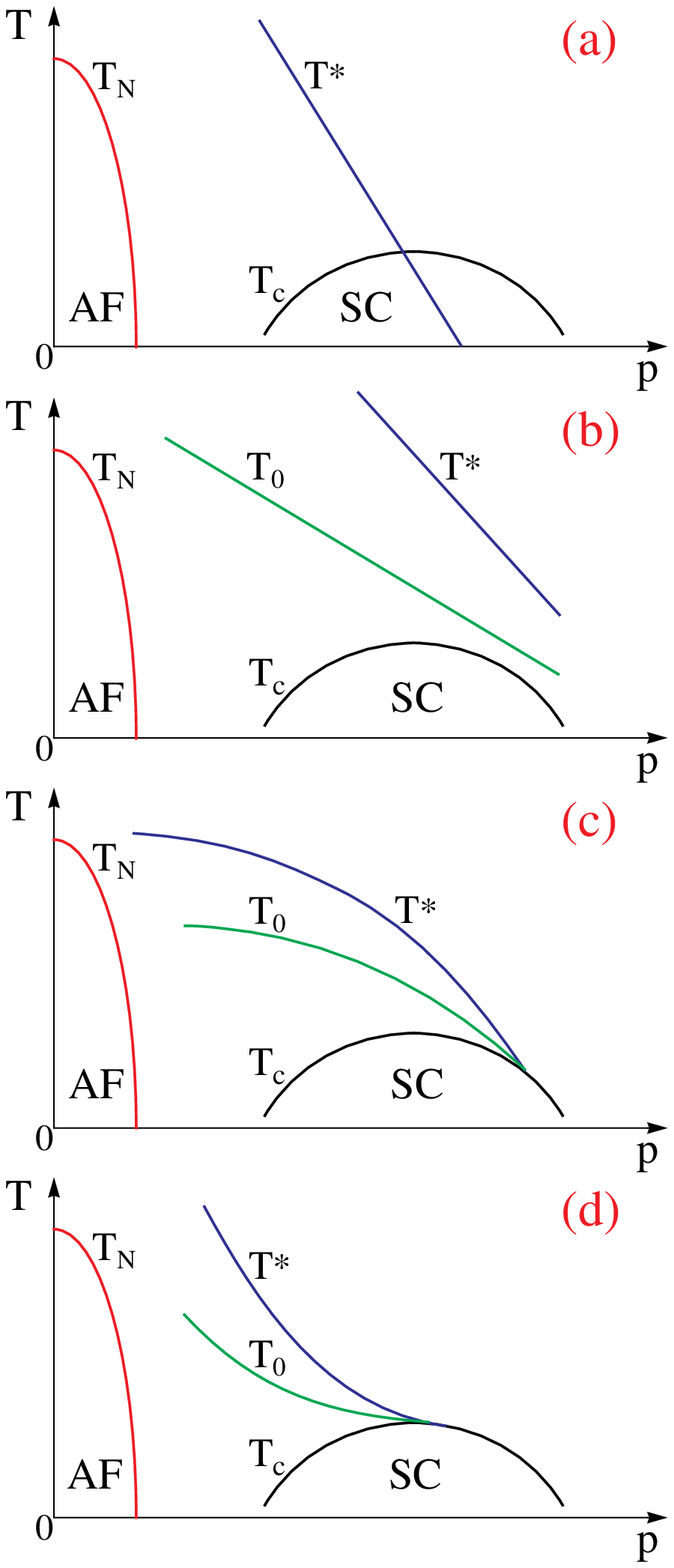}}
\vspace{2mm} 
\caption{Different phase diagrams of superconducting cuprates, which can be 
found in the literature. $T_N$ is the N\protect{\'e}el temperature; $T_c$ is 
the critical temperature; $T^{\ast}$ is the pseudogap temperature; and $T_0$ 
is the pairing temperature (AF = antiferromagnetic; SC = superconducting).}
\label{fig1} 
\end{figure} 
However, here the common agreement ends. Further there is a polarization of 
opinions based either on different measurements or on different theories. First, 
there is no yet consensus on the presence of preformed Cooper pairs above 
$T_c$ in cuprates. Secondly, there is no agreement on the exact doping 
dependence of temperature (energy) scale $T^{\ast}$. 

The question concerning the presence of preformed Cooper pairs above $T_c$ 
in cuprates is not the main topic of this paper, therefore we are not going to 
discuss it here. There are enough evidence in the literature for the 
presence of  preformed Cooper pairs above $T_c$ in cuprates (see
references in [2] and [3]).  The purpose of this paper is to discuss the
second issue raised above, namely,  the doping dependences of different
temperature/energy scales in hole-doped  cuprates and their origins.  
Hole-doped cuprates have a very rich phase diagram as a function of doping. 
This is due to a charge inhomogeneity in CuO$_2$ planes: the charge 
distribution in CuO$_2$ planes is inhomogeneous both on a nanoscale and 
macroscale. Depending on the doping level, the CuO$_2$ planes comprise 
at least four different types of clusters having a size of a few nanometers. 
Each type of clusters has its own features. As a consequence, the common 
phase diagram consists of, at least, seven different energy/temperature 
scales. As an example, we consider a phase diagram 
of Bi$_{2}$Sr$_{2}$CaCu$_{2}$O$_{8+x}$ (Bi2212). 

\section{Different phase diagrams} 

In figure 1, the first two phase diagrams are based on experimental data: 
the phase diagram shown in figure 1(a) is mainly inferred from nuclear 
magnetic resonance, specific heat and resistivity measurements (see 
references in [3]). The phase diagram in figure 1(b) is based on tunneling 
and angle-resolved photoemission (ARPES) measurements. 
In figure 1(b), the pseudogap temperature scale $T^{\ast}$ is inferred 
from bias positions of humps in tunneling and ARPES spectra, while the 
positions of main peaks in these spectra determine the pairing 
temperature/energy scale $T_0$ corresponding to the formation of 
incoherent Cooper pairs.  The phase diagrams in 
figures 1(c) and 1(d) are purely theoretical and based on the idea of the 
existence of two types of superconductivity: an unconventional type in 
underdoped cuprates and a BCS type in overdoped cuprates. They emerge 
near the optimally doped region of superconducting phase. 
 
Leaving the phase diagrams shown in figures 1(c) and 1(d) for further 
theoretical considerations, we focus our attention exclusively on 
experimental data presented in figures 1(a) and 1(b). The question is why do 
different experimental techniques give different energy/temperature 
scales? This question is directly related to the understanding of 
normal-state properties of cuprates. 
\begin{figure}[h]
\leftskip-10pt
\epsfxsize=1.0\columnwidth
\centerline{\epsffile{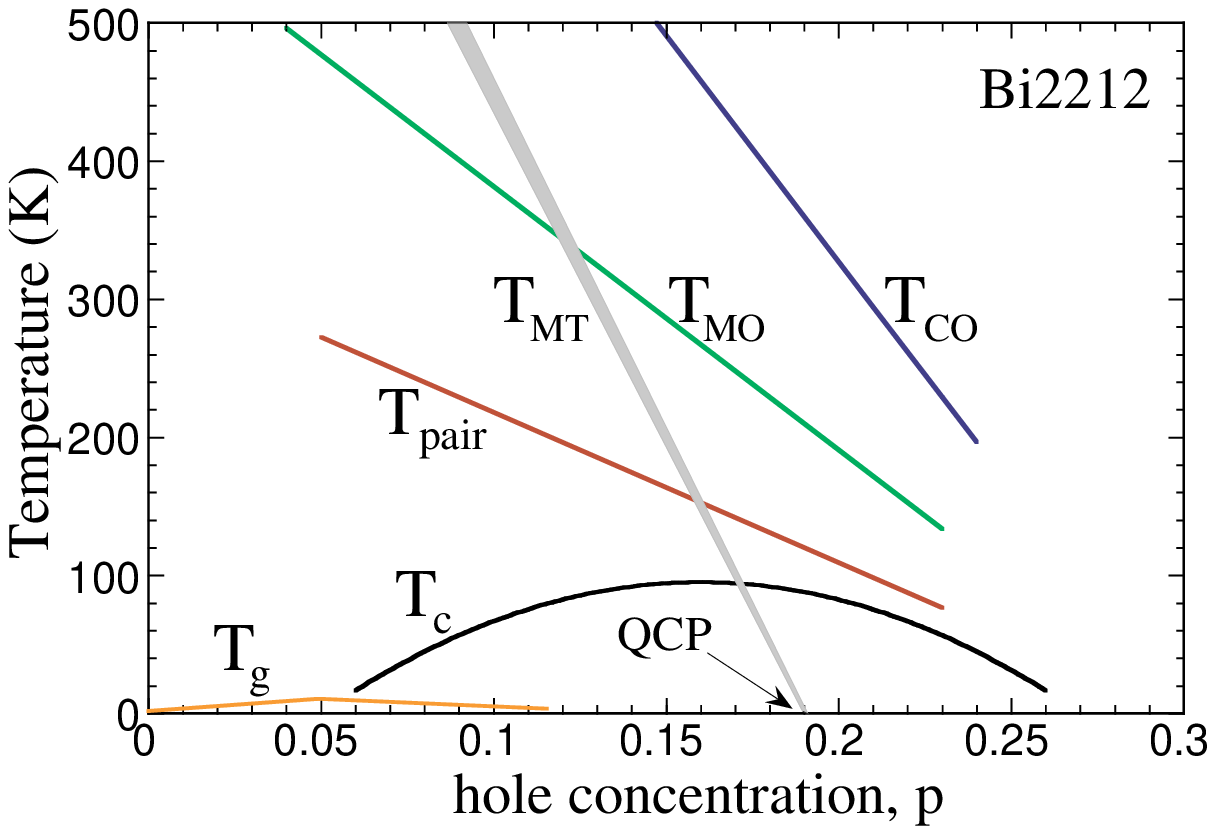}}
\vspace{2mm} 
\caption{Phase diagram of Bi2212 [3]. $T_c$ is the critical temperature; 
$T_{pair}$ is the pairing temperature; $T_{MT}$ is the magnetic-transition 
temperature; $T_{MO}$ is the magnetic-ordering temperature; and $T_{CO}$ 
is the charge-ordering temperature (for more details see text). The 
commensurate antiferromagnetic phase at low doping is not shown. The 
spin-glass temperature scale $T_g$ is shown schematically (QCP = 
quantum critical point).}
\label{fig2}
\end{figure}

\section{Phase diagram of Bi2212} 

Figure 2 shows an idealized phase diagram of Bi2212 taken from [3], in
which the  spin-glass temperature scale $T_g$ is added. This
temperature/energy scale  was recently observed in Bi2212 by muon spin
relaxation measurements [4].  Figure 2 shows six different
temperature/energy scales in Bi2212. The  seventh temperature scale---the
commensurate antiferromagnetic phase at low doping level---is not shown
because it was not yet observed in Bi2212:  There is a technical problem
to synthesize large-size good-quality  {\em undoped} single crystals of 
Bi2212, which are necessary for neutron scattering measurements. In fact, 
the commensurate antiferromagnetic phase is not the primary focus of 
this paper. It is worth noting that the phase diagram of Bi2212 in figure 
2 is {\bf not} universal but reflects main features of the physics involved in 
all cuprates. First, the temperature scales in figure 2 will be given 
explicitly, and then we discuss their origins. 

The superconducting phase appears at a critical temperature $T_c$ 
shown in figure 2. In Bi2212, the doping dependence $T_c (p)$ can be 
expressed [5] as
\begin{equation} 
T_c (p) \simeq T_{c,max} [1 - 82.6(p - 0.16)^2]. 
\end{equation} 
For Bi2212, $T_{c,max}$ = 95 K. The superconducting phase is 
approximately located between $p$ = 0.05 and 0.27, having the maximum 
critical temperature $T_{c,max}$ in the middle, thus at $p \simeq$ 0.16. The 
corresponding phase-coherence energy scale $\Delta _{c}$ is proportional 
to $T_c$ as $2 \Delta _{c} = \Lambda k_B T_c$, where $k_B$ is the
Boltzmann  constant. In different cuprates, the coefficient $\Lambda$ is
slightly different: 
$\Lambda \simeq 5.4$ in Bi2212; $\Lambda \simeq$ 5.1 in YBCO; and 
$\Lambda \simeq 5.9$ in  Tl$_{2}$Ba$_{2}$CuO$_{6}$ (Tl2201) [3]. 

In the literature, there are ample evidence for the existence of the so-called
quantum  critical point in cuprates (see references in [3]). In a quantum
critical point  located at/near absolute zero, the magnetic order is about to
form  or to disappear. In figure 2, the quantum critical point in Bi2212 exists
at 
$p \simeq$ 0.19. At low temperature, superconductivity in cuprates is most 
robust at this doping level $p$ = 0.19, and not at $p$ = 0.16 (see below why). 

In figure 2, the temperature scale $T_{MT}$ starts/ends in the 
quantum critical point (MT = Magnetic Transition). Then, it is more or less 
obvious that this temperature/energy scale has a magnetic origin. 
The temperature scale $T_{MT}$ is analogous with a magnetic transition 
temperature of long-range antiferromagnetic phase in heavy fermions [3]. 
The doping dependence $T_{MT} (p)$ can be expressed as  
\begin{equation} 
T_{MT} (p) \simeq T_{MT, 0} \times \left[ 1 - \frac{p}{0.19} \right],   
\end{equation} 
where $T_{MT, 0} =$ 970--990 K (see references in [3]). As was 
mentioned above, this energy scale is inferred from resistivity, nuclear 
magnetic resonance and specific heat measurements. Magnetic fluctuations 
are strong along the transition temperature $T_{MT}(p)$. 

In figure 2, the energy scale with a characteristic temperature $T_{CO}$ 
corresponds to a charge ordering (CO) in shape of quasi-one-dimensional 
stripes. A structural phase transition precedes the charge ordering. The 
doping dependence $T_{CO}(p)$ can approximately be expressed as follows [3] 
\begin{equation} 
T_{CO} (p) \simeq 980 \times \left[ 1 - \frac{p}{0.3} \right]
\quad\mbox{(Kelvins).}\quad 
\end{equation} 
The corresponding charge gap $\Delta _{cg}$ observed in tunneling 
and ARPES spectra depends on hole concentration as 
\begin{equation} 
\Delta _{cg} (p) \simeq 251 \times \left[ 1 - \frac{p}{0.3} \right]
\quad\mbox{(meV).}\quad
\end{equation} 

In cuprates, manganites and nickelates, the charge ordering always precedes 
a magnetic ordering (MO) [3]. In figure 2, the energy scale with 
a characteristic temperature $T_{MO}$ corresponds to an antiferromagnetic 
ordering occurring in insulating stripes which separate charge stripes. 
This magnetic ordering stabilizes the charge order which becomes 
long-ranged at $T_{MO}$. The doping dependence $T_{MO}(p)$ can be 
expressed as follows 
\begin{equation} 
T_{MO} (p) \simeq 566 \times \left[1 - \frac{p}{0.3} \right]
\quad\mbox{(Kelvins).}\quad   
\end{equation} 
The dynamical two-dimensional magnetic stripes can be considered 
as a local memory effect of the commensurate antiferromagnetic phase. 

In figure 2, the temperature scale $T_{pair}$ corresponds to the formation of 
Cooper pairs, the doping dependence of which can be expressed as  
\begin{equation} 
T_{pair} (p) \simeq \frac{T_{CO} (p)}{3} = \frac{980}{3} \times \left[ 1 - 
\frac{p}{0.3} \right]
\quad\mbox{(Kelvins).}\quad   
\end{equation} 
The corresponding pairing energy scale $\Delta _{p}$ depends on 
doping level as follows 
\begin{equation} 
\Delta _{pair} (p) \simeq \frac{\Delta _{cg} (p)}{3} = 
\frac{251}{3} \times \left[ 1 - \frac{p}{0.3} \right] 
\quad\mbox{(meV).}\quad 
\end{equation} 
The pairing gap manifests itself in tunneling and ARPES measurements. 
The extensions of three dependences $T_{CO}(p)$, $T_{MO}(p)$ and 
$T_{pair}(p)$ cut the horizontal axis {\em approximately} in one point, at 
$p$ = 0.3. 

In figure 2, the spin-glass temperature scale $T_g$ is shown schematically. 
For example, the temperature scale $T_g$ in LSCO linearly increases as the 
doping starts to increase from zero, reaches its maximum when it crosses 
a N\'eel temperature scale $T_N$, and then decreases to zero as the doping 
increases. In Bi2212, $T_g \simeq$ 8 K at $p$ = 0.05, and $T_g$ = 0 
somewhere at $p \simeq$ 0.15 [4]. 

It is worth noting that, in first approximation in Bi2212, the two 
energy scales with characteristic temperatures $T_{CO}$ 
and $T_{MT}$ intersect the vertical axis in one point, at $T \approx$ 980 K. 
Some discussion on this issue can be found elsewhere [3]. 

It is necessary to mention that the above formula are approximate. 
To understand why, consider just one temperature/energy scale---the 
charge ordering scale $T_{CO}$. From acoustic measurements, it is well known  
that any structural phase transition has a hysteresis on lowering and on 
increasing the temperature. Since in cuprates, a structural phase transition 
precedes the charge ordering, its hysteresis may affect the charge ordering 
temperature scale. As a consequence, the above expression for 
$T_{CO}(p)$ is most likely approximate. Secondly, the real doping dependences of 
temperature scales $T_{MT}$, $T_{CO}$, $T_{MO}$ and $T_{pair}$, most likely, are
not linear but quasi-linear. 

\section{Charge inhomogeneities: phase separation} 

Let us go back to the question raised above: Why do different experimental 
techniques give different energy/temperature scales? 
As a matter of fact, this question is related to another question, 
namely, how is it possible that there are several magnetic
temperature/energy  scales in one phase diagram (see figure 2)? 
\begin{figure}[h]
\leftskip-10pt
\epsfxsize=0.9\columnwidth
\centerline{\epsffile{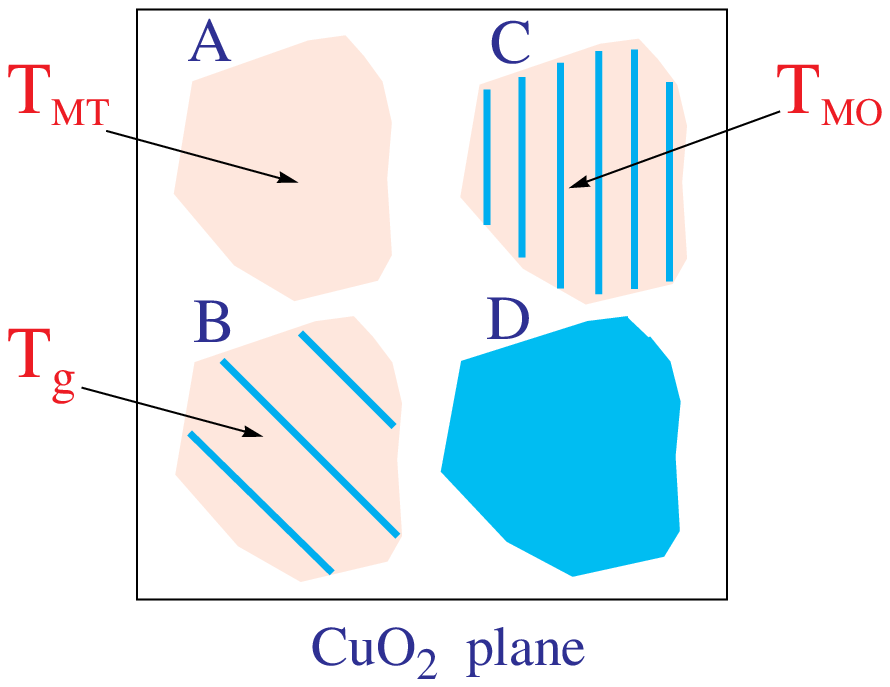}}
\vspace{2mm} 
\caption{Schematic representation of four types of clusters existing in 
CuO$_2$ planes at different dopings. Cluster A includes a virgin 
antiferromagnetic phase. Cluster B includes diagonal charge stripes 
and a magnetic ordering between the stripes. The stripes are 
schematically shown by the lines. Cluster C includes vertical charge 
stripes and a magnetic ordering between the stripes. In cluster D, holes 
are distributed homogeneously (the Fermi sea). The magnetic ordering in
clusters  A, B and C occurs at $T_{MT}$, $T_g$ and $T_{MO}$, respectively.}
\label{fig3}
\end{figure} 

Consider CuO$_2$ planes at different doping levels. In undoped region 
($p <$ 0.05), the doped holes gather into clusters having a size of a few 
nanometers. These clusters are embedded in an antiferromagnetic 
background where the long-range antiferromagnetic order is preserved. In 
the clusters, the holes are not distributed homogeneously, but they form 
quasi-one-dimensional stripes along the diagonal direction relative to the 
Cu--O--Cu bonds [6]. In figure 3, this type of clusters is marked by 
B. The spin glass appears in these clusters at $T_g$ [6]. 

In underdoped region, thus above $p =$ 0.05, the charge stripes in
most clusters are rotated by 45$^{\circ}$ relative to those in
undoped region. Thus, they now run along the Cu--O--Cu bonds, and the
distance between stripes becomes smaller than that between diagonal
stripes. However, in a small fraction of clusters, the charge stripes
remain  diagonally oriented. This explains the existence of spin
glass in the  superconducting phase. The clusters with vertically
(horizontally) oriented  charge stripes are schematically shown in figure
3 by C. Thus, in deep underdoped region, the two types of
clusters---with vertical stripe order  and with diagonal stripe
order---are embedded in an antiferromagnetic  background. 
Therefore, the long-range antiferromagnetic phase which is usually located at 
low doping level extends to higher dopings. The doped holes are sucked out by 
the clusters. As a consequence, the N\'eel temperature scale $T_N$ literally 
transforms into the temperature scale $T_{MT}$ shown in figure 2. In a sense, 
the temperature scale $T_{MT}$ is an echo of the $T_N$ scale. 
\begin{figure}[h]
\leftskip-10pt
\epsfxsize=0.8\columnwidth
\centerline{\epsffile{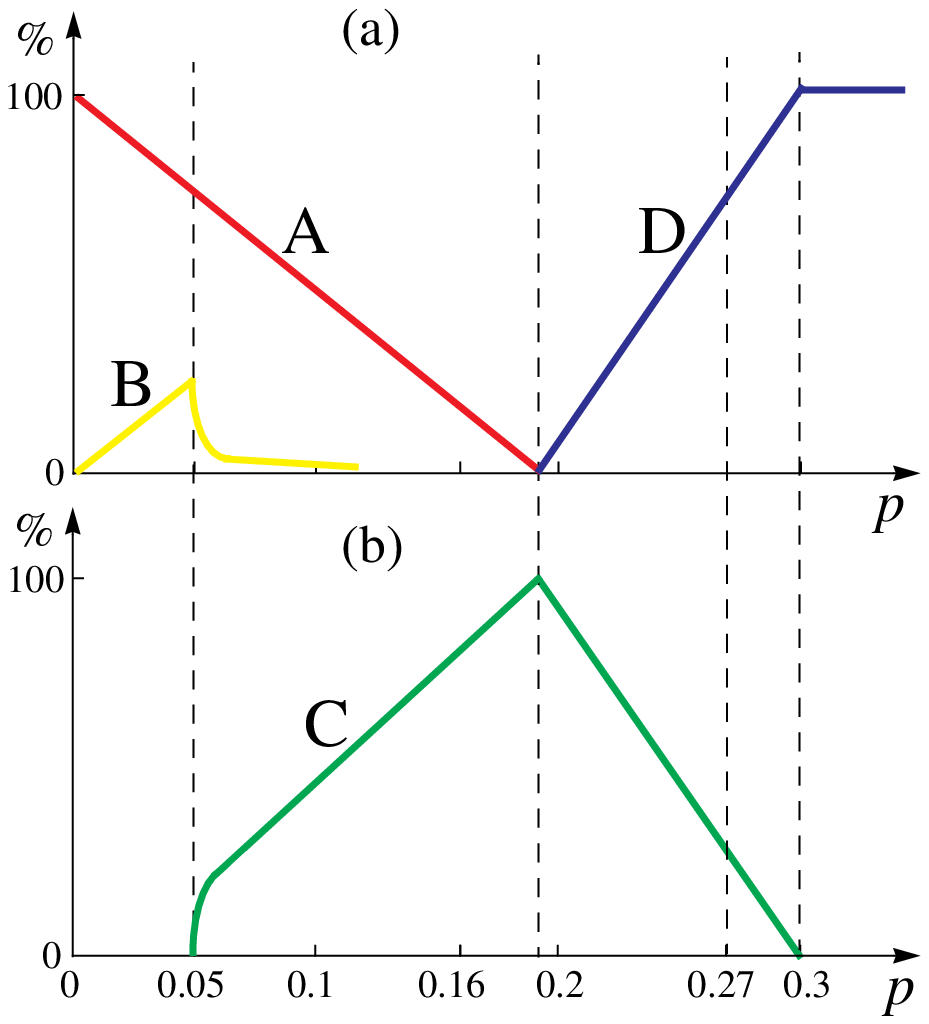}}
\vspace{2mm} 
\caption{(a) Fractions of clusters (phases) A, B and D in figure 3 in a 
CuO$_2$ plane as a function of doping. (b) Fraction of clusters (phase) C. 
Both plots are shown schematically.}
\label{fig4}
\end{figure}  

Near optimally doped region, the picture is different: now clusters 
with intact antiferromagnetic order are embedded in a charge-stripe 
background. In figure 3, the antiferromagnetic clusters are shown by A. 
At $p$ = 0.19, this type of clusters completely vanishes, as schematically 
shown in figure 4(a). At the same time, the fraction of clusters with 
vertical charge stripes is maximal at $p$ = 0.19, as shown in 
figure 4(b). Since superconductivity in cuprates occurs in these clusters, 
this is the reason why it is most robust at $p$ = 0.19, and not at $p$ = 0.16. 

The three temperature scales, $T_{MT}$, $T_g$ and $T_{MO}$ shown in figure 
2, originate from a magnetic ordering in different clusters---A, B and C in 
figure 3, respectively. Since different experimental techniques are sensitive 
to different types of correlations, and have different resolutions and different 
characteristic times, it is then obvious why there is a discrepancy among 
phase diagrams presented in the literature. 

Finally, let us consider the deep overdoped region. In deep overdoped 
region, new doped holes gather between stripes, forming clusters with Fermi 
sea, in which the hole distribution is more or less homogeneous. The 
Fermi-sea clusters, shown schematically in figure 3 by D, are embedded in 
a charge-stripe background. Since, the superconducting phase vanishes at 
$p$ = 0.27, it is commonly assumed that above this doping level, the hole 
distribution becomes homogeneous in CuO$_2$ planes. In fact, this 
assumption may be not true: The three temperature scales $T_{CO}$, $T_{MO}$ 
and $T_{pair}$ in figure 2 originate from different types of ordering in  
clusters with vertical charge stripes.  Since these three temperature
scales cut the horizontal axis approximately at $p$ = 0.3, as shown in
figure 2, it is possible that between $p$ = 0.27 and  0.3, the hole
distribution in CuO$_2$ planes is not completely homogeneous.  The
clusters with vertical charge stripes may exist between $p$ = 0.27  and
0.3, being embedded in a Fermi-sea background. In this case, the hole 
distribution in CuO$_2$ planes is homogeneous only above $p$ = 0.3, as 
schematically shown in figure 4(a). After all, this is not very important 
for understanding the mechanism of high-$T_c$ superconductivity. 

\section{Stripe phase} 

Since superconductivity in cuprates occurs in clusters with vertical 
charge stripes, it is most important to understand the physics of this phase. 
The size of clusters with vertical charge stripes depends on the 
doping level: In underdoped region, the size of these clusters is 
about 30 \AA \ [7]. Then, it is obvious that the length of charge stripes is
of the same magnitude. In overdoped region, the length of charge
stripes increases up to 100 \AA \ [8]. Therefore, the length of charge stripes
also depends on the doping level. 

Since charge stripes in cuprates are very dynamical, one may wonder how 
it is possible to observe charge-stripe domains by tunneling [7, 8]. First of all, 
there are two types of dynamics involved: the fluctuation of charge stripes 
inside clusters, and the clusters fluctuate as whole. Superconductivity 
most likely requires the former but not the latter. Secondly, the charge stripes 
can slow down in one separate cluster, this will not affect 
superconductivity locally {\bf if} the charge stripes in the neighboring clusters 
go on fluctuating [3]. Generally speaking, the charge stripes can be pinned by 
an impurity, by a lattice defect, or by the surface.  

It is commonly assumed that the charge-stripe orientation in adjacent 
CuO$_2$ layers is alternately rotated by 90$^{\circ}$. In fact, this 
assumption is not necessary: Depending on doping level, the clusters 
with {\em vertical} charge stripes may coexist in the same CuO$_2$ layer 
with clusters having {\em horizontal} charge stripes [3]. 

\section{Summary} 

We discussed here a phase diagram of hole-doped cuprates and the origins of 
different temperature/energy scales. 
Hole-doped cuprates have a very rich phase diagram as a function of doping. 
This is due to a charge inhomogeneity in CuO$_2$ planes: the charge 
distribution in CuO$_2$ planes is inhomogeneous both on a nanoscale and 
macroscale. Depending on the doping level, the CuO$_2$ planes comprise 
at least four different types of clusters having a size of a few nanometers. 
Each type of clusters has its own features. As a consequence, the common 
phase diagram consists of, at least, seven different energy/temperature 
scales. As an example, we considered a phase diagram 
of Bi$_{2}$Sr$_{2}$CaCu$_{2}$O$_{8+x}$.

\end{document}